\documentclass[pre,aps,superscriptaddress,twocolumn]{revtex4-2}
\usepackage{graphicx}
\usepackage[caption=false]{subfig}
\usepackage{epstopdf}
\usepackage{amsmath, physics}
\usepackage{amssymb}
\usepackage[usenames,dvipsnames]{xcolor}

\usepackage[normalem]{ulem}

\newcommand{\rev}[1]{{#1}}

\newcommand*{\kt}{k_\text{B}T}

\newcommand*{\Es}{\epsilon_\text{s}}
\newcommand*{\Eb}{\epsilon_\text{b}}
\newcommand*{\Kb}{\kappa_\text{b}}
\newcommand*{\KbStar}{\kappa_\text{b}^*}

\newcommand{\cBar}{\bar{c}}
\newcommand{\Nspecies}{N_\text{species}}
\newcommand{\Nedges}{N_\text{edge-types}}

\newcommand{\Tmag}{T_\text{magic}}

\newcommand{\Na}{n_\text{A}}

\begin{document}

\title{Magic sizes enable minimal-complexity, high-fidelity assembly of programmable shells}

\author{Botond Tyukodi}
\email{botond.tyukodi@ubbcluj.ro}
\affiliation{Department of Physics, Babe\c{s}-Bolyai University, 400084 Cluj-Napoca, Romania}
\affiliation{Centre International de Formation et de Recherche Avanc\'{e}es en Physique, 077125 Bucharest-M\={a}gurele, Romania}
\affiliation{Martin Fisher School of Physics, Brandeis University, Waltham, Massachusetts 02454, USA}
\author{Fernando Caballero}
\affiliation{Martin Fisher School of Physics, Brandeis University, Waltham, Massachusetts 02454, USA}
\author{Daichi Hayakawa}
\affiliation{Martin Fisher School of Physics, Brandeis University, Waltham, Massachusetts 02454, USA}
\author{Douglas M. Hall}
\affiliation{Department of Polymer Science and Engineering,University of Massachusetts, Amherst, Massachusetts 01003, USA}
\author{W. Benjamin Rogers}
\affiliation{Martin Fisher School of Physics, Brandeis University, Waltham, Massachusetts 02454, USA}
\author{Gregory M. Grason}
\email{grason@umass.edu}
\affiliation{Department of Polymer Science and Engineering,University of Massachusetts, Amherst, Massachusetts 01003, USA}
\author{Michael F. Hagan}
\email{hagan@brandeis.edu}
\affiliation{Martin Fisher School of Physics, Brandeis University, Waltham, Massachusetts 02454, USA}

\pacs{--}

\begin{abstract}
Recent advances in synthetic methods enable designing subunits that self-assemble into structures with precise, finite sizes and well-defined architectures, but yields are frequently suppressed by the formation of off-target metastable structures. Increasing the complexity (the number of distinct subunit types) can inhibit off-target structures, but leads to slower kinetics and higher synthesis costs. Here, we study icosahedral shells formed of programmable triangular subunits as a model system, and identify design principles that produce the highest target yield at the lowest complexity. We use a symmetry-based construction to create a range of design complexities, starting from the maximal symmetry Caspar-Klug assembly up to the fully addressable, zero-symmetry assembly. Kinetic Monte Carlo simulations reveal that the most prominent defects leading to off-target assemblies are disclinations at sites of rotational symmetry. We derive symmetry-based rules for identifying the optimal (lowest-complexity, highest-symmetry) design that inhibits these disclinations, leading to robust, high-fidelity assembly of targets with arbitrarily large, yet precise, finite sizes. The optimal complexity varies non-monotonically with target size, with `magic' sizes appearing for high-symmetry designs in which symmetry axes do not intersect vertices of the triangular net.  The optimal designs at magic sizes require $12$ times fewer inequivalent interaction-types than the (minimal symmetry) fully addressable construction, which greatly reduces the timescale and experimental cost required to achieve high fidelity assembly of large targets. \rev{This symmetry-based principle for pruning off-target assembly generalizes to diverse architectures with different topologies.}

\end{abstract}

\date{\today}

\maketitle

The self-limiting assembly of subunits into structures with precise, finite sizes and well-defined architectures underlies many of the essential functions in cells and the pathogens that infect them (e.g. virus capsids
\cite{Zlotnick2011, Mateu2013, Bruinsma2015, Perlmutter2015, Hagan2016, Twarock2018, Zandi2020, Hagan2021}, bacterial microcompartments \cite{Schmid2006, Iancu2007, Kerfeld2010, Rae2013, Chowdhury2014, Kerfeld2016}, and cellular protein shells \cite{Sutter2008, Pfeifer2012, Nott2015, Zaslavsky2018}). 
Inspired by these biological assemblies, recent advances in DNA origami, protein design, supramolecular chemistry, and patchy-colloidal particles have enabled designing synthetic subunits  that undergo self-limited assembly into analogous architectures, such as quasi-spherical shells and cylindrical tubes with programmable, finite sizes up to thousands of subunits.~\cite{Rothemund2004, Benson2015, Sigl2021, Videbaek2022, Hayakawa2022, Wei2024, Videbaek2024, Wagenbauer2017, Divine2021, Butterfield2017, Bale2016, Hsia2016, King2012, Dowling2023, Malay2019, Ren2019, Laniado2021, McConnell2020, McMullen2022}. Despite these spectacular successes, it remains a challenge to achieve the high \emph{fidelity} (yield of a target with precisely controlled size and structure) required for many functions. 

In particular, the fidelity falls dramatically with increasing target size because thermal bending fluctuations of the subunit-subunit angles within the growing assemblage lead to competing off-pathway structures \cite{Sigl2021, Hagan2021}. This effect becomes more pronounced with increasing size, unless reinforced by the presence of a curved scaffold ~\cite{Li2018b}. 
While these fluctuations can be suppressed by increasing the geometric specificity, i.e., the relative cost of binding with off-target angles and distances \cite{Hagan2006, Wilber2007}, achievable stiffnesses are \rev{inevitably} limited by material properties and physical dimensions of subunits. Thus, it is a generic feature of self-assembly that accessible, high-yield target sizes are limited by fluctuations.   
Alternatively, fidelity can be increased by increasing the design complexity (number of specifically interacting subunit species) to make off-pathway structures inaccessible \cite{Bupathy2022, Goodrich2024, King2024, Videbaek2024}, but synthesis costs and assembly timescales increase with the complexity ~\cite{Videbaek2024, Whitelam2009,Whitelam2015} and become impractical for large targets. Thus, there is a critical need for new strategies to develop designs that minimize the complexity required for high-fidelity assembly.

\begin{figure*} 
\centering
\includegraphics[width=.99\textwidth]{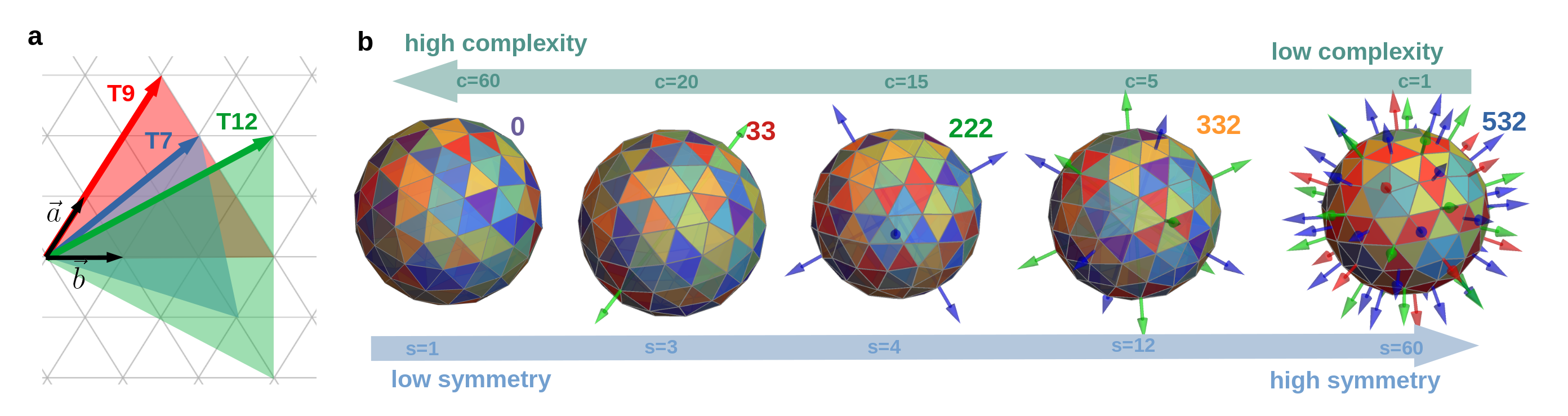}
\caption{
(a) Construction of icosahedron facets of increasing size ($T$ number). (b) Sequence of  $T{=}9$ programmable shells, from left to right showing progressively {\it increasing symmetry} ($s=N_\mathrm{operations}$) and  correspondingly {\it decreasing complexity} ($c= \Nedges / T$). Structures are labeled according to their orbifold symmetry denoting the nature of rotational symmetries in the structure (see SI sections II-IV).  For clarity, we illustrate the 5-fold (red), 3-fold (green), and 2-fold (bue) rotational symmetries.}
\label{fig:symmetry_construction}
\end{figure*}

\begin{figure}
\begin{center}
\includegraphics[width=\columnwidth]{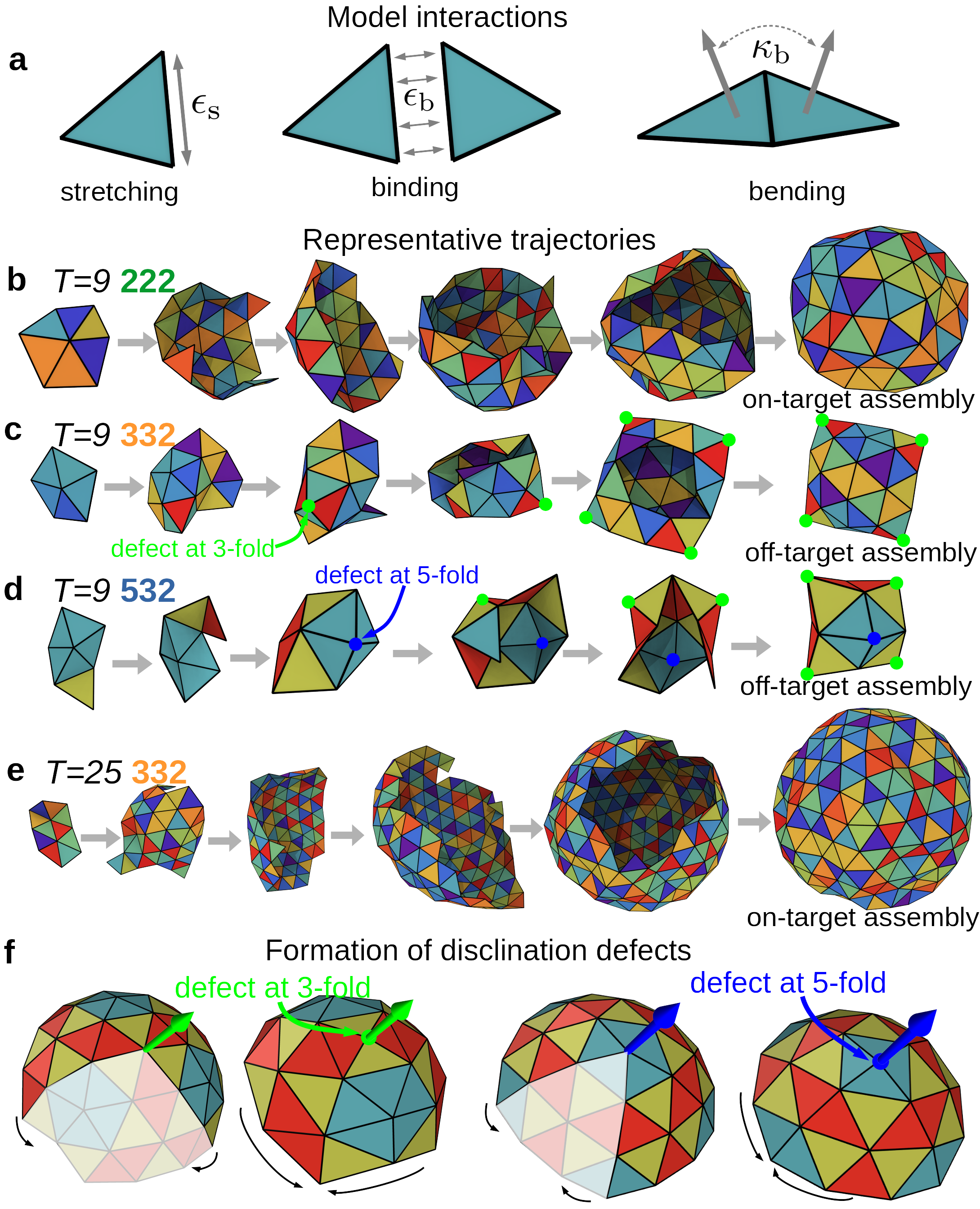}
\caption{
(a) Subunit edges have a stretching modulus $\Es$, complementary edges bind with affinity $\Eb$, and adjacent faces have a preferred dihedral angle $\theta_0$, and a harmonic bending energy modulus $\Kb$. 
(b-d) Snapshots from trajectories for a $T{=}9$ target with  $\Kb=\kt/2$ and $\theta_0 = 0.234 $ for the designs: (a) 222, (b) 332, (c) 532 (CK). Only the 222 design assembles successfully. The 332 is driven off-pathway by the formation of $+2\pi/3$ disclinations around the 3-fold symmetry axis. The 532  design mis-assembles due to  $+2\pi/3$ disclinations around the 3-fold axis and either $+2\pi/5$ or $+4\pi/5$ disclinations around the 5-fold axis. (e) Snapshots from an on-target assembly trajectory for $T{=}25$ with $\Kb=\kt$ and $\theta_0 = 0.134$ for the 332 design. See Videos 1-4. (f) The Volterra construction illustrating possible disclinations at a $p$-fold site, visualized by the removal of a wedge of angle $2\pi/p$ (highlighted in white) and reclosing around along compatible edges, distorting according to black arrows. 
}
\label{fig:disclinations}
\end{center}
\end{figure}

In this letter, we study icosahedral shells as a model system to identify the principles that underlie a `Pareto optimal' set of target structures that achieve high fidelity at the minimal possible design complexity. The principles derive from symmetry rules for obstructing disclinations, the most prominent class of defects, which lead to metastable but long-lived off-pathway structures that suppress target yields. Remarkably, in contrast to lower complexity designs (Ref. ~\cite{Li2018b}), the optimal complexity designs achieve high-fidelity assembly of large shells without a scaffold, which can contain 500 or more subunits (limited only by our computational power), Fig.~\ref{fig:disclinations}b and SI Fig. 1). 
Significantly, we identify a spectrum of `magic' shell sizes that minimize the optimal complexity, enabling high fidelity assembly with 12 times fewer interaction types than perfect specificity (i.e. a fully `addressable' assembly in which every subunit is a distinct species with a unique `address' in the target \cite{Jacobs2015, Jacobs2015a, Jacobs2016, Zeravcic2014, Madge2017, Sajfutdinow2018, Hedges2014, Whitelam2015a}). \rev{Our results, and the symmetry-based design rules they illuminate, can be generalized to a broad class of self-closing architectures with different topologies.}

\textit{Designing icosahedral shells from low complexity to fully addressable.}
Developed to rationalize and categorize the icosahedral structures of natural viral capsid shells, Caspar and Klug (CK) devised a symmetry-based theory \cite{Caspar1962, Johnson1997} to systematically construct icosahedral shells with increasing sizes, by subdividing each of the 20 facets of the base icosahedron into increasing numbers of subunits occupying the minimum number of unique symmetry environments (Fig.~\ref{fig:symmetry_construction}a).  Here, we start with CK shells as the limiting case of maximal-economy programmable shells.   Connecting any two points of a triangular lattice gives an edge of a base facet of desired size; repeating this base facet constructs the whole icosahedron. The base edge vector is given by $\vb{T}=h \vb{a} + k\vb{b}$ where $\vb{a}, \vb{b}$ are the lattice base vectors with lengths in units of the triangular lattice spacing $l_0$ and $h, k$ are integers. The number of triangles in a base facet is $T=|\vb{T}|=h^2+k^2+hk$. Thus, $T$ is the minimal number of subunit species required to form an icosahedron with a given size.

We choose triangular subunits for our model icosahedra as the simplest realization of the CK construction, motivated by DNA origami icosahedral shell designs \cite{Sigl2021, Wei2024} and natural viruses whose capsids assemble from protein trimers (e.g. \cite{Medrano2016}). Icosahedra feature three kinds of rotational symmetry axes (we do not consider mirror symmetry): 2-fold,  3-fold, and 5-fold axes on the base edges, facet centers, and base vertices respectively.
CK assigns the same species for subunits in equivalent local symmetry environments, resulting in $\lceil T/3 \rceil $ species  and $T$ distinct edge-types for triangular subunits (where each edge corresponds to an individual protein in the CK construction).  The preferred edge-lengths and inter-triangle dihedral angles must be slightly adjusted away from equilateral to avoid elastic costs of geometric frustration (see SI section X).  Although we focus on icosahedra assembled from triangular subunits, our results will be qualitatively similar for other subunit shapes when adapted for the topology of their tiling in spherical nets.  

To investigate how assembly depends on complexity and symmetry, we construct  designs that systematically vary between the minimal complexity needed to avoid geometric frustration (i.e. CK rules) and the fully addressable case, in which every subunit is distinct. 
Adopting the orbifold notation for spherical symmetry patterns \cite{Conway2008}, we start with the fully addressable 0 pattern, which has no symmetries as all triangles are distinct.  Then we consider assemblies with a subset of the full symmetries of the icosahedron, corresponding to subgroup combinations of 5-, 3- and 2-fold axes. Fig. \ref{fig:symmetry_construction}b shows a sequence of increasing symmetry for $T{=}9$ nets: 0 (fully addressable), 33, 222, 332 and 532 (CK) (see SI sections II-IV for the remaining 22, 55, 322 and 522 symmetries and further explanation).

Including symmetry axes reduces the number of distinct species compared to the fully addressable 0 structures, since operating the rotational symmetry elements on the structure maps equivalent triangles (and edges) onto their multiple locations (or `addresses') in the assembly.  Hence, increasing the number of symmetries in a design increases the copy number of a particular triangle (subunit species) in the target structure, i.e., decreases the number of distinct subunit- and edge-types in the target structure, making it less complex. We define \textit{complexity} as the number of distinct edge-types normalized by $T$: $c=\Nedges / T$ (Fig. \ref{fig:symmetry_construction}b). \rev{Our conclusions hold for other complexity measures (see SI).} Similarly, we quantify symmetry $s=N_\mathrm{operations}$ as the number of symmetry operations for which facets are equivalent.  


\textit{Results.}
To investigate the assembly dynamics, we perform kinetic Monte Carlo (KMC) simulations using a model adapted from previous works \cite{Rotskoff2018, Panahandeh2018, Wagner2015, Panahandeh2020, Li2018b, Panahandeh2022, Duque2023a, Tyukodi2022, Videbaek2022, Fang2022, Mohajerani2022a} (Fig. \ref{fig:disclinations}a). We represent \rev{assembling structures as elastic shells}, within which subunits are triangles with spring constant $\Es$ along each edge. Each subunit has a species (distinguished by color in snapshots) and each edge has an edge-type. Edges that are complementary according to the target designs bind to each other with affinity $\Eb$, and non-complementary edges cannot bind. The ground-state curvature is programmed by a preferred dihedral angle $\theta_0$ between adjacent face normals, with a harmonic penalty for angular deviations $0.5 \Kb (\theta-\theta_0)^2$ with bending modulus $\Kb$. 
\rev{This model reduces to elasticity theory of spherical shells in the continuum limit \cite{Li2018b}.} 
The bending and edge stretching moduli define the {\it geometric specificity} of inter-subunit binding,  i.e., the relative cost of binding with off-target angles and distances (Fig. \ref{fig:disclinations}a.).  \rev{We simulate a single shell growing in the presence of a reservoir of free subunits at constant chemical potentials, set to maintain concentrations of species $c_i$ in the reservoir that reflect their stoichiometries in the target shell, while keeping the mean concentration $\cBar=\sum_{i=1}^{\Nspecies} c_i / \Nspecies$   the same for all designs. The KMC algorithm includes 13 moves  that account for
subunit association/dissociation and structural relaxation of intermediates.} \rev{We show in SI section VIII that the results are not sensitive to changes in these relative rates. More significantly, we tested the central results of the KMC simulations against particle-based molecular dynamics simulations.  The results show strong qualitative agreement with the KMC results, indicating that realistic assembly pathways at fixed total concentration exhibit the same sensitivity to the symmetry (see End Matter).} 
 Further model details are in SI sections VI-VII and Refs.~\cite{Tyukodi2022,Duque2023a}.
For results in the main text, we set the stretching modulus near the rigid limit, $\Es=300 \kt/l_0^2$, and the binding affinity to $\Eb=-7 \kt$ with the mean concentration $\cBar=6.7\times 10^{-3} l_0^{-3}$ to ensure that nucleation times are not prohibitively large but that subunit binding is sufficiently reversible to permit annealing of mis-bound subunits \cite{Hagan2014,Ceres2002}. Our conclusions hold for other binding affinities (SI section V).



We performed simulations for sizes from $T{=}1-16$ over a range of geometric specificity (bending modulus $\Kb$) values.
Fig. \ref{fig:disclinations} shows typical trajectories of three designs for a $T{=}9$ target at moderate geometric specificity $\Kb=\kt/2$. We observe that 222 assembles on-pathway  and closes into the $T{=}9$ target, notwithstanding bending fluctuations (Fig.~\ref{fig:disclinations}a). In contrast, the 332 and 532 (CK) designs fail  by closing at structures that are smaller than the target (Fig.~\ref{fig:disclinations}c,d).  
The predominant mechanism for misassembly is the formation of a generalization of disclination defects that form around symmetric vertices.  As in standard disclinations in crystals \cite{Seung1988}, these are defects where the preferred $p$-fold symmetry of vertex is disrupted, for example by closing with $p\pm 1$ triangles around a given vertex, as shown schematically in Fig. \ref{fig:disclinations}e for 5- and 3-fold axes of the 532 structures.  \rev{Such defects lead to metastable, but extremely long-lived, off-target structures, which are generally smaller than the target geometry but still close while satisfying edge-matching rules required by the specific interactions.}

\begin{figure}
\begin{center}
\includegraphics[width=\columnwidth]{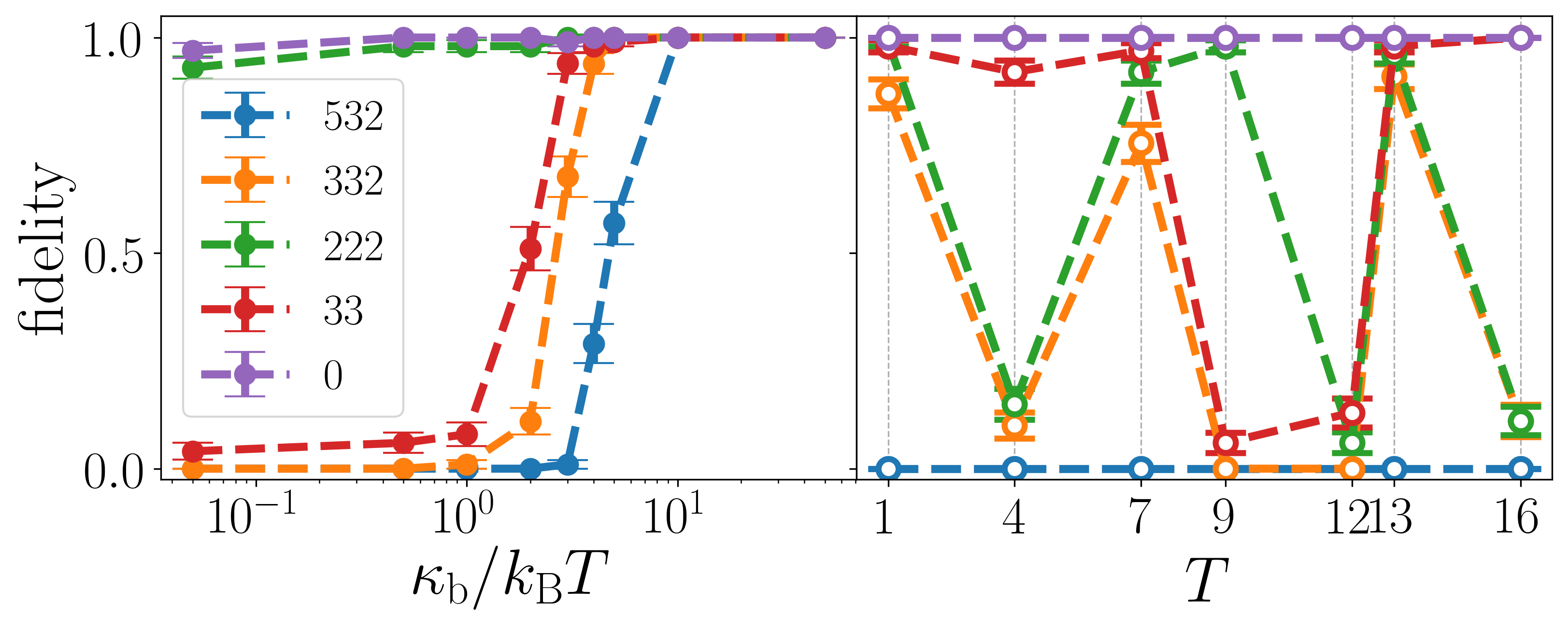}
\caption{
(a) Fidelities of $T{=}9$ targets for the five designs as a function of bending modulus $\Kb$  (geometric specificity). 
(b) Fidelities of each design as a function of shell size for $\Kb=\kt/2$. Fidelities were computed from 100 independent simulations at each parameter set. 
Error bars are smaller than the symbol size for many points.
}
\label{fig:yield}
\end{center}
\end{figure}

Our simulations show that designs that inhibit formation of disclinations lead to remarkably robust high-fidelity assembly.
Fig. \ref{fig:yield}a shows the \emph{fidelity} (defined as the fraction of nucleated trajectories that result in the target structure) for each design of a  $T{=}9$ target structure as a function of $\Kb$. Above a threshold $\KbStar \approx 1 - 10 \kt$, we observe near 100\% fidelity for all designs because the high bending modulus prohibits disclinations. 
However, below this threshold the fidelities for the 532 and then 332 and 33 fall to nearly 0, while 222 and 0 (fully addressable) remain high even at $\Kb=0.05 \kt$. Except near $\KbStar$, the fidelities are nearly independent of $\Kb$.
The fidelities decrease mainly due to closed off-target, non-icosahedral structures facilitated by disclinations at $p$-vertices, (we mostly observe ``preclosure'' to lower-coordinated vertices than target values, causing the assembly to close before reaching the target size). 

Notably, the fidelity is not always monotonic with complexity (and hence symmetry order), because it also depends on the locations of symmetry axes (as we explain below). \rev{For example, 222 performs much better than 33 despite having lower complexity. 
}

To further understand the interplay between complexity, symmetry, and fidelity, Fig. \ref{fig:yield}b shows the fidelities for each design as a function of $T$.
As expected, the fully-addressable design 0 always gives the highest fidelity. However, at most (but not all) $T$-numbers, one or more lower complexity designs perform nearly as well. We define the \textit{optimal complexity} as the lowest complexity design that results in a fidelity of $\ge 75\%$.
Notably, the optimal complexity varies non-monotonically with target size and, as for $T{=}9$, larger complexity is not necessarily better. For example, 332 is optimal for $T{=}1,7,13$;  222 for $T{=}9$; 33 for $T{=}4$; but for $T{=}12$ the fully-addressable design is optimal. The CK design is suboptimal for all sizes and always gives $\approx 0$ fidelity for $\Kb < \KbStar$.

\begin{figure}
\begin{center}
\includegraphics[width=\columnwidth]{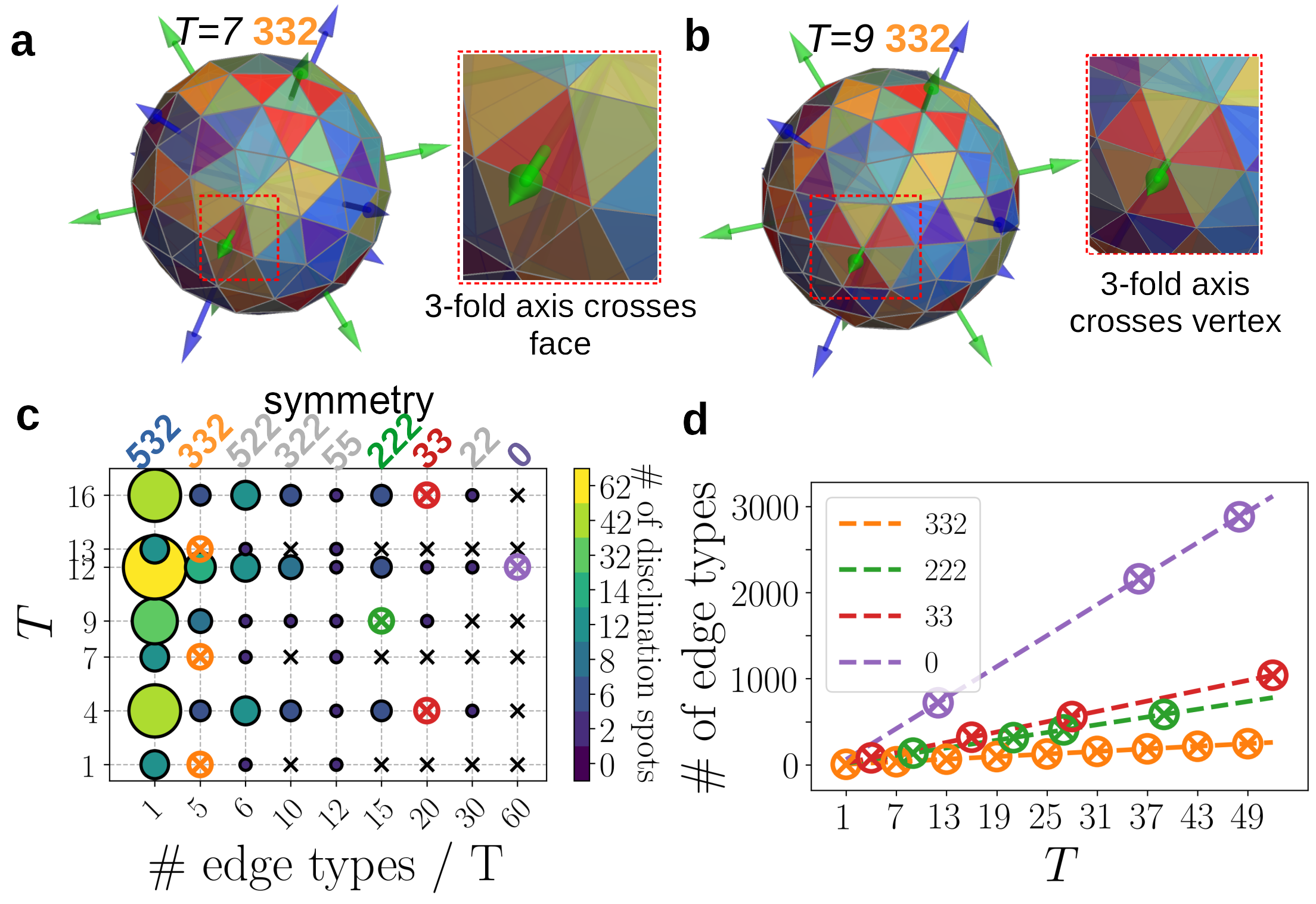}
\caption{Magic $T$ numbers enable extremely economic high-fidelity assembly. 
(a-b) Locations of the 3-fold symmetry axes for the 332 designs with $T{=}7$ and $T{=}9$. For $T{=}7$ the 3-fold axis (in green) crosses through a \emph{facet} (red), corresponding to the subunit center, and thus does not allow for disclinations. For $T{=}9$ the 3-fold axis crosses through a hexameric \emph{vertex}, which allows disclinations. 
 (c) The number of disclination points (locations where a symmetry axis crosses a vertex)  for indicated complexities and $T$. Markers are colored and sized by the number of disclination points, and for each $T$ number the `x' symbol marks the optimal complexity at which no disclination points remain. Magic $T$ numbers correspond to sizes at which the 2-fold and two 3-fold symmetry axes all cross through facets for the lowest complexity design without a 5-fold axis, 332. 
 (d) The number of distinct edge-types for the optimal complexity design as a function of shell size. The dashed lines show the upper- and lower-bounds, in which the optimal design corresponds to 332 or 0 respectively. The x-axis labels show the magic $T$-numbers.  
}
\label{fig:magicT}
\end{center}
\end{figure}

\rev{The nonmonotonic dependency of the fidelity on the complexity and target size can be 
can be explained by the location of the points of rotational symmetry relative to points where subunits bind (i.e. vertices). Fig. \ref{fig:magicT}(a-b) show the symmetry axes of $T{=}7$ and $T{=}9$ for the 332 design. While the 2-fold axes cross edges in both designs, the 3-fold axes cross \emph{facets} (i.e., subunit centers) for $T{=}7$ and \emph{vertices} for $T{=}9$. This key difference allows disclinations to readily form at vertices (by addition/removal of $2 \pi/3$ wedges) during assembly of $T{=}9$, but not for $T{=}7$}. 
Following this reasoning, we hypothesize that \emph{the optimal complexity corresponds to the lowest complexity (highest symmetry) design  that does not have symmetry axes crossing a vertex}. For example, for $T{=}7$, all of 0, 33, 22, and 332 have no symmetry axes crossing vertices and exhibit robust assembly, so 332 is the optimal complexity. For $T{=}9$, only 0 and 222 lack symmetry axes crossing vertices, so 222 is the optimal complexity. Since 5-fold axes necessarily cross through pentameric vertices, any design with a 5-fold symmetry axis allows disclinations for all $T$. This feature explains why the CK (532) and 55 designs always exhibit low fidelities for low $\Kb$.

Thus, while rotational symmetries increase design economy, they can also induce the formation of off-target assemblies when the symmetry points correspond to vertices in the target net. In support of this interpretation, the threshold specificity $\KbStar$ increases with the number of possible disclination points (see Fig. \ref{fig:yield}a and SI Fig. 13).


\textit{Magic $T$ numbers.}
The optimal complexity design can be predicted from geometric considerations: whenever $(2h+k)/3 \in \mathbb{Z}$ and $(k-h)/3 \in \mathbb{Z}$, the 3-fold axes cross at vertices, allowing $\pm 2 \pi/3$ defects. Similarly, whenever $h/2 \in \mathbb{Z}$ and  $k/2 \in \mathbb{Z}$, the 2-fold axes cross a vertex, allowing $\pm \pi$ defects.  Thus, $T$-numbers at which both these conditions are satisfied allow disclinations if they possess any 2-, 3- or 5-fold symmetries, and the fully addressable design will be optimal. Sizes that satisfy one condition will have an intermediate optimal complexity. Sizes that satisfy neither condition correspond to the \textit{magic} $T$-numbers, which assemble robustly at any complexity that does not include a symmetric 5-fold axis (Fig. \ref{fig:magicT}c and SI Fig. 7).  The periodicity of these conditions gives rise to a pattern of optimal complexities ranging between 332 and 0, and a series of magic $T$-numbers, $\{\Tmag=1+6n \, | \, \forall \, n\in \mathbb{Z} \}$.
Fig. \ref{fig:magicT}d illustrates the number of  distinct edge-types required for the optimal complexity design as a function of shell size.  The lower bound corresponds to the  magic $T$-numbers, where the 332 designs assemble robustly with 12 times fewer edge-types than the fully addressable case. 

\textit{Conclusions.} We have demonstrated principles to identify the minimal complexity subunit design that achieves high-fidelity assembly of icosahedral shells with arbitrarily large target sizes. The designs inhibit formation of the primary class of defects that facilitate \rev{metastable off-pathway assembly --- disclinations at $p$-vertices --- and thereby account for kinetics as well as thermodynamics to maximize fidelity. In contrast, many existing algorithms to identify optimal designs for icosahedral shells do not a-priori account for defects that form during dynamics  \cite{Bohlin2023, Pinto2023, Pinto2024, Romano2020, Mart2021, Osat2024, Evans2024, Pinto2024a, Pandey2011, Dodd2018, Curatolo2023}.}
Any design, including the fully addressable case, requires moderate binding affinities and subunit concentrations to avoid kinetic traps
\cite{Zlotnick1999,Endres2002,Zlotnick2003,Ceres2002,Hagan2006,Jack2007,Rapaport2008,Whitelam2009,Nguyen2007,Wilber2007,Wilber2009,Hagan2011,Cheng2012,Hagan2014,Whitelam2015,Perlmutter2015,Panahandeh2020,Asor2020,Zandi2020,Qian2023}.
However, the optimal designs achieve nearly 100\% fidelities over a broad range of parameter values, even for vanishing geometric specificity ($\Kb \lesssim 0.05 \kt$). Furthermore, for sub-optimal designs, the threshold specificity $\KbStar$ required for high fidelities increases with the number of possible disclination points. 
While the class of defects we identify is relevant for icosahedral shells, the same principles can be applied to nets with other symmetry elements (e.g. including 4-fold rotations).  \rev{For example, analogous types of defects suppressing target yields were identified simulations of negative-curvature triply periodic frameworks \cite{Duque2023a} and  in experiments and simulations on assembly of toroidal structures \cite{saha2025}. This suggests that our strategy can be applied to diverse assembly architectures. }

Finally, recent advances in DNA origami~\cite{Sigl2021} and protein design~\cite{Dowling2023} have enabled assembly of CK shells up to $T=100$, but with low or unmeasured yields. We anticipate that applying our design principles to these technologies can enable high-fidelity assembly of icosahedra and a wide variety of other geometries with precisely controlled sizes and architectures. Moreover, the ability of optimal designs to assemble at low geometric specificity will enable a broad array of synthesis techniques with less precision than DNA origami and protein engineering.

The data files and code used to generate the results will be publicly distributed at the time of publication.

-----------------------
------------------------

\begin{acknowledgments}
 This work was supported by the NSF through DMR 2309635 (MFH, BT) and the Brandeis Center for Bioinspired Soft Materials, an NSF MRSEC (DMR-2011846) (MFH, DH, WBR, BT, DMH, GMG). This project has received funding from the European Union’s Horizon 2020 research and innovation programme under the Marie Skłodowska-Curie grant agreement No 01026118, PNRR-I8/C9-CF105 under contract 760099 and project number PN-IV-P2-2.1-TE-2023-0558, within PNCDI IV from the Romanian Ministry of Research, Innovation, and Digitization, CNCS-UEFISCDI(BT). Computing resources were provided by the National Energy Research Scientific Computing Center (NERSC), a Department of Energy Office of Science User Facility (award BES-ERCAP0026774); the NSF XSEDE allocation TG-MCB090163 (Expanse and Anvil); and the Brandeis HPCC which is partially supported by the NSF through DMR-MRSEC 2011846 and OAC-1920147.
\end{acknowledgments}

\section*{End Matter}

\begin{figure}
    \centering
    \includegraphics[width=0.95\linewidth]
    {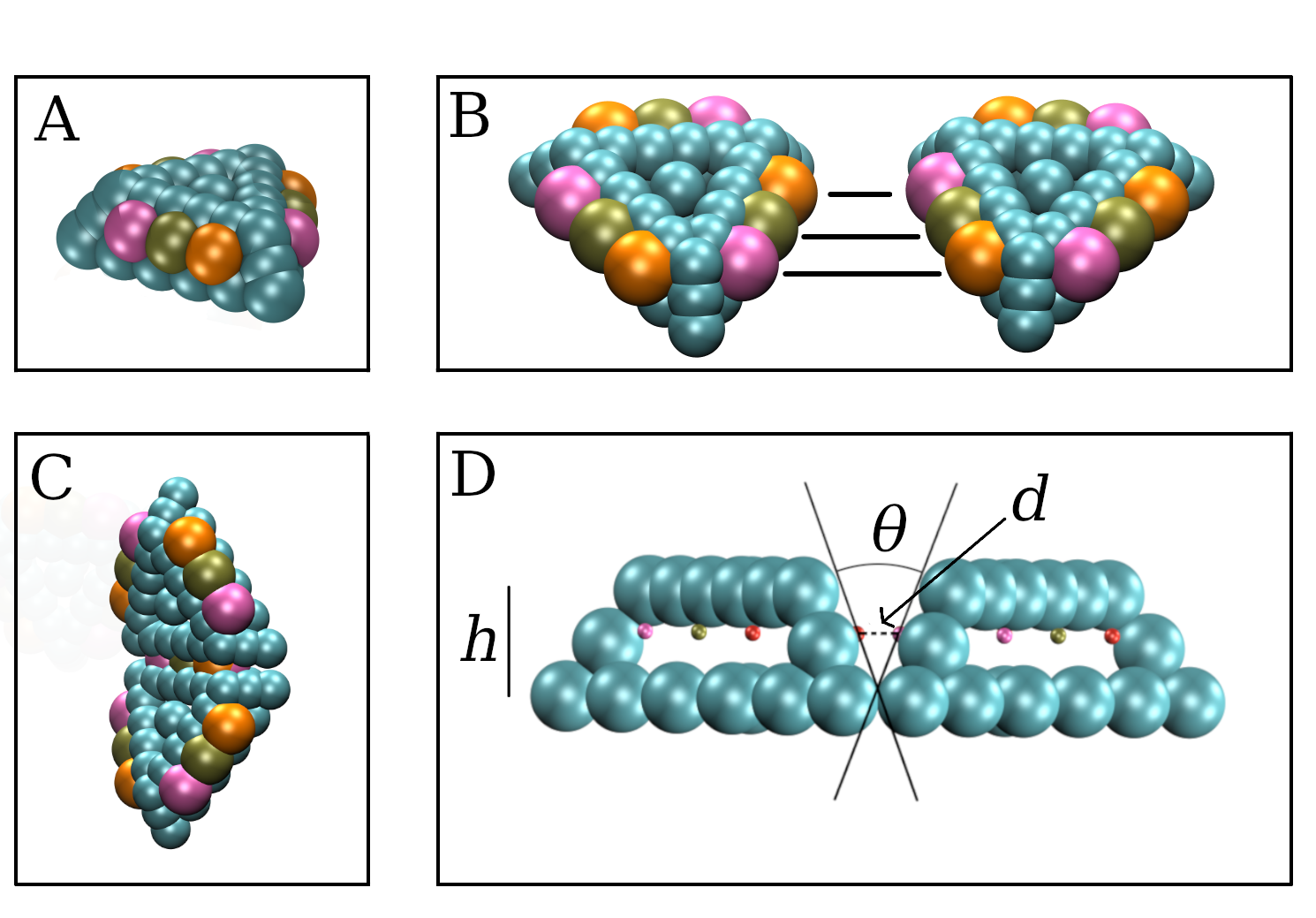}
    \caption{(A) A subunit, with 35 excluder pseudoatoms (responsible for excluded volume) in blue and $\Na=3$ attractor pseudoatoms in pink, green and orange. (B) Attractive interactions, with pairs of complementary pseudoatoms connected by solid black lines. (C) Minimum energy configuration for a dimer. (D) Illustration of the calculation of the bending modulus. For an angle $\theta$, the distance between the attractors is $d = 2h\sin(\theta/2)$, where $h$ is the subunit height. Attractive pseudoatoms have been drawn smaller for clarity. }
    \label{fig:model}
\end{figure}

To test that our conclusions are not affected by the approximations required for the KMC model, we have developed an analogous particle-based model (Fig.~\ref{fig:model}) and performed Brownian dynamics (BD) simulations in the NVT ensemble. Thus, in contrast to the KMC model, multiple structures can assemble at the same time at fixed total concentration (allowing, in principle, for monomer starvation kinetic traps), we explicitly simulate particle diffusion, transition rates emerge naturally from the dynamics, and there are no subunit exchanges with a bath.  
We designed the subunits to closely replicate the triangular subunits of the KMC model. We control the  subunit-type specific interactions to match the constructions described in the main text (Fig. 1 main text), and we vary the angular potential between bound subunits to quantitatively control the bending modulus. The model is described in detail in SI section XI. Given the significant increase in computational cost  of BD simulations compared to the KMC model, we focus on a critical comparison of $T=1$ capsids.

\begin{figure}
    \centering
    \includegraphics[width=0.95\linewidth]{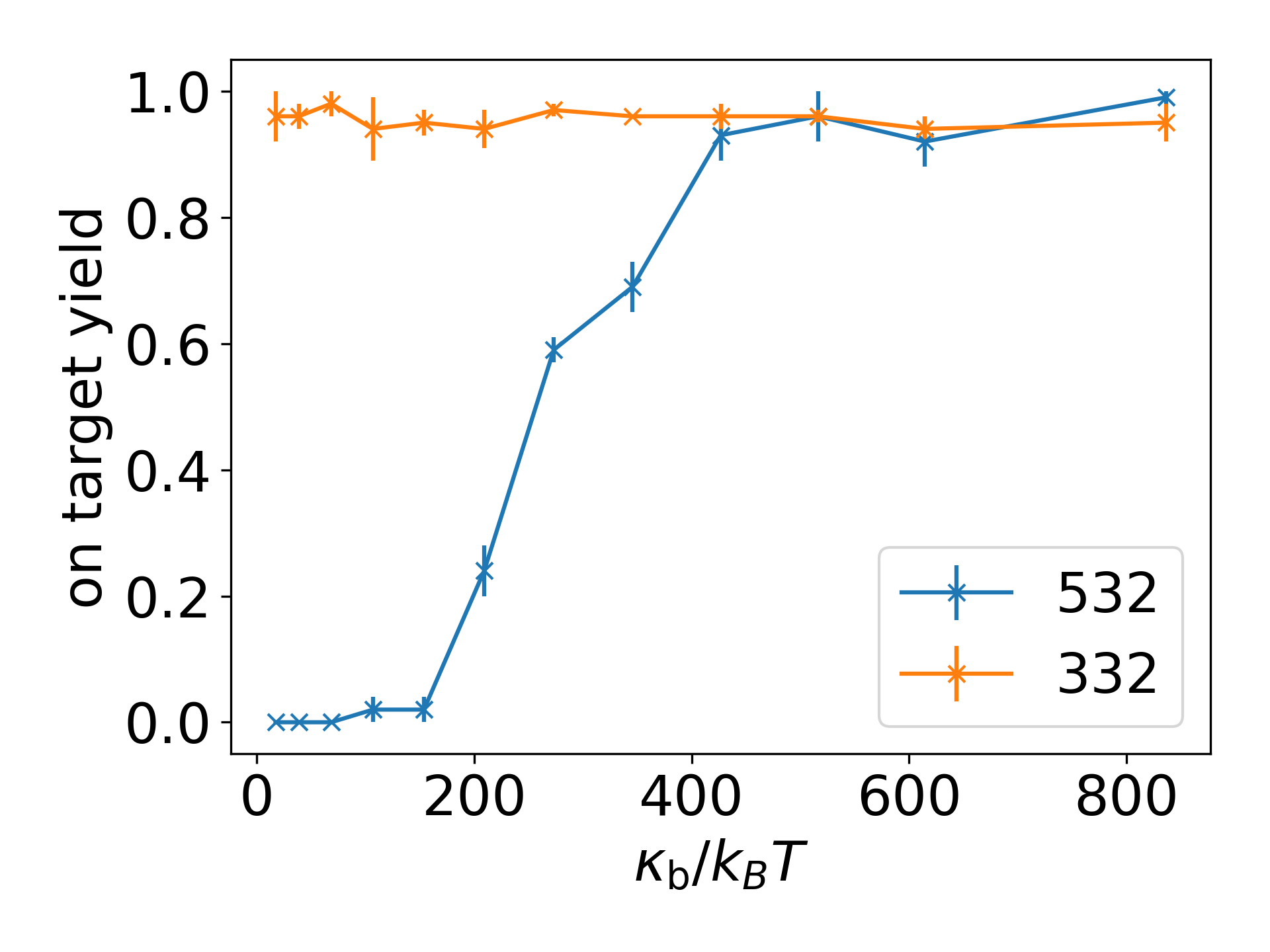}
    \caption{Yield as a function of bending modulus $\Kb$ for the 532 and 332 designs. Errorbars are twice the standard mean error over 5 independent trajectories.}
    \label{fig:results}
\end{figure}

\begin{figure}
    \centering
    \includegraphics[width=0.5\linewidth]{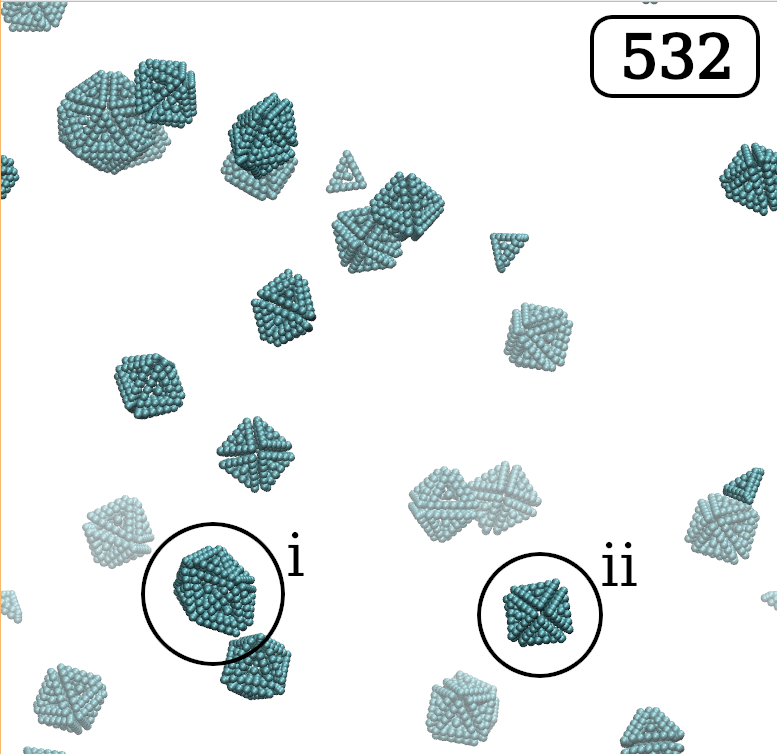}\includegraphics[width=0.5\linewidth]{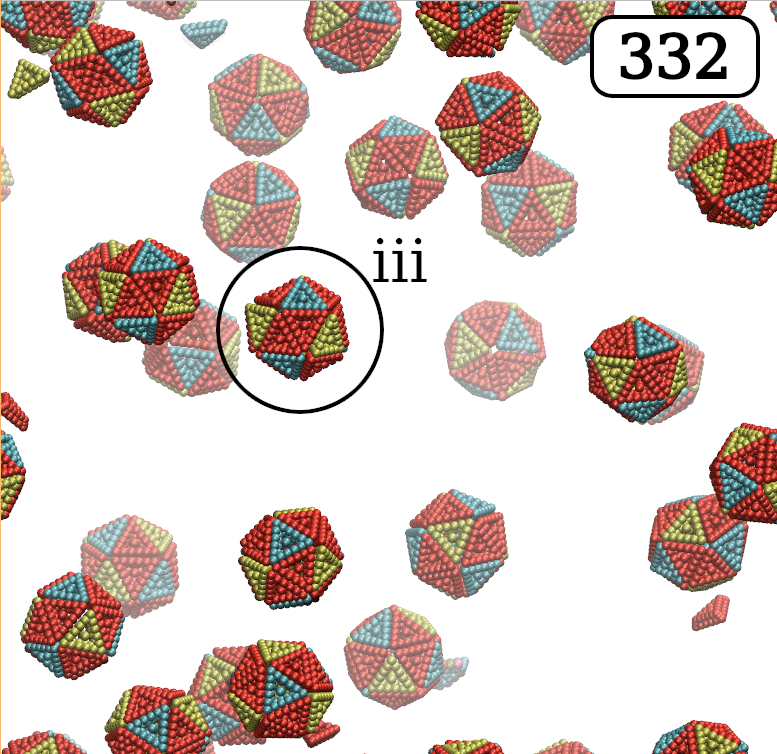}
    
    \includegraphics[width=0.9\linewidth]{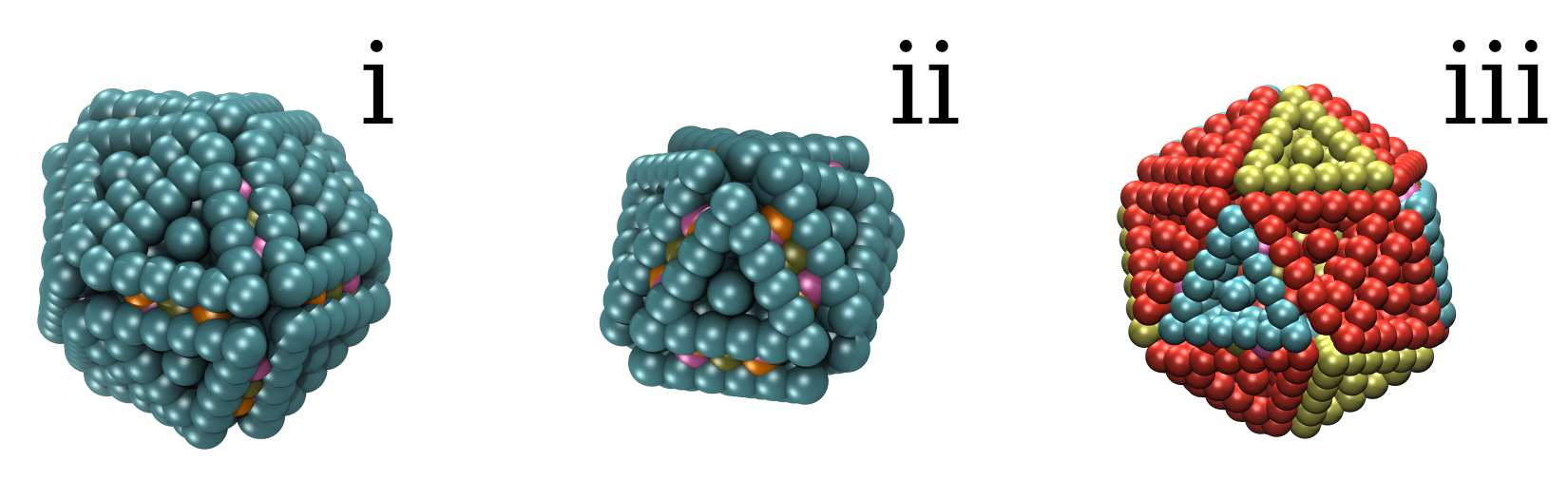}
    \caption{Top: Representative snapshots for $\Kb=100\kt$, showing that the low yields in the 532 design arise from disclinations that lead to aberrant capsids. Bottom: close-up views of two malformed capsids for the 532 design, and a capsid for the 332 design. See corresponding Videos 5 and 6.}
    \label{fig:defects}
\end{figure}

Fig.~\ref{fig:results} compares the yields for the 532 (CK limit) and 332 (optimal) designs as a function of $\Kb$, and Fig.~\ref{fig:defects} shows representative snapshots of assembly dynamics for each case. Since we are using the NVT ensemble, we define the \emph{yield} as the fraction of subunits in well-formed capsids. The dependence of yields on complexity is strikingly similar to the KMC results. While the 332 design yields approach 100\% across the range of simulated $\Kb$ (up to 98.7\%), there is a threshold $\KbStar\approx 300 \kt$ below which yields for the 532 design decrease to nearly zero. Note that achieving exactly 100\% yields is not possible in the NVT ensemble since there must be a finite concentration of free subunits at equilibrium. Most significantly, the reduced yields for 532 arise from exactly the same class of disclinations predicted by the KMC model (Fig.~\ref{fig:defects}), yielding smaller off-target closed structures, such as octahedra.

Interestingly,  $\KbStar$ is an order of magnitude larger than the corresponding value for the KMC results ($\KbStar \cong 10 \kt$ for $T=1$). We attribute this to differences in other aspects of the geometric specificity between the two models. Reducing the bending modulus in the BD model also lowers the moduli for the other modes of angular fluctuations (torsion and twist). In contrast, these fluctuations are absent in the KMC model since bound subunits share the same edge. Similarly, bound subunits in the BD model experience translational displacements between edges that are absent in the KMC model.  Despite these differences, and the corresponding quantitative difference between the threshold bending modulus value, the dependence of yields on bending modulus and complexity is remarkably similar between the two models. 

\textit{Discussion.}
The results from the BD simulations in Fig.~\ref{fig:results} and \ref{fig:defects} strongly support the conclusions of the main text, as the BD model relaxes the key approximations required for the KMC model. The model can be readily extended to the full range of designs and capsid sizes, but simulation times will become less tractable for large sizes. 

We note that this class of KMC models has been previously used to describe diverse geometries such as capsids, tubules, and negative-curvature triply-periodic frameworks \cite{Rotskoff2018, Panahandeh2018, Wagner2015, Panahandeh2020, Li2018b, Panahandeh2022,  Tyukodi2022, Videbaek2022, Fang2022, Mohajerani2022a,Duque2023a}, but has not been previously compared against particle-based simulations. Thus, our comparison between the KMC and BD simulations also supports the observations of these previous works and future investigations with the KMC model.

\bibliographystyle{aipauth4-1}
\bibliography{specificity_paper}
\bibliographystyle{unsrt}
\end{document}